\documentclass[fleqn,amsmath]{revtex4}
\usepackage{graphicx}
\usepackage{longtable}

\begin{document}

\title{Precision spectroscopy of the molecular ion {\rm HD}$^+$: control of Zeeman shifts
}

\author{Dimitar Bakalov}
\address{Institute for Nuclear Research and Nuclear Energy,
Tsarigradsko chauss\'{e}e 72, Sofia 1784, Bulgaria}

\author{Vladimir Korobov}
\address{Bogolyubov Laboratory of Theoretical Physics, JINR, Dubna 141980, Russia}

\author{Stephan Schiller}
\address{Institut f\"ur Experimentalphysik, Heinrich-Heine-Universit\"at
D\"usseldorf, D-40225 D\"usseldorf, Germany}

\begin{abstract}
Precision spectroscopy on cold molecules can potentially enable
novel tests of fundamental laws of physics and alternative
determination of some fundamental constants. Realizing this
potential requires a thorough understanding of the systematic
effects that shift the energy levels of molecules. We have
performed a complete {\em ab initio} calculation of the magnetic
field effects for a particular system, the heteronuclear molecular
hydrogen ion HD$^+$. Different spectroscopic schemes have been
considered, and several transitions, all accessible by modern
radiation sources and exhibiting well controllable or negligible
Zeeman shift, have been found to exist. Thus, HD$^+$ is a
 candidate for the determination of the ratio of
electron-to-nuclear reduced mass, and for tests of its
time-independence.
\end{abstract}

\maketitle

 Cold molecules have recently been proposed as novel
 systems for precision measurements related to fundamental
 aspects of physics, such as the measurement of the electron-to-nuclear mass ratio,
 its possible time-variability,
 tests of Lorentz Invariance, tests of QED, and parity violation.
 Several of these are based on precision
 spectroscopy  \cite{Froehlich,kor&schil,schiller08,Mueller,mkajita,flambaum,deMille,ye},
 where frequencies of ro-vibrational transitions exhibiting high quality factors must be
 measured. As the
 tests mentioned above have already been performed with very high precision using
 various atomic systems, studies on molecular systems must be
 conceived in ways that have the potential of surpassing
 atomic tests. This implies that molecular systems must be
 experimentally accessible and that systematic shifts of the transition frequencies must be
 sufficiently small so as to guarantee the desired spectroscopic accuracy.

 We focus on a particular diatomic molecule, HD$^+$. Because of its
 relative simplicity it can be
 analyzed with high-precision {\em ab initio} methods.
 QED calculations of ro-vibrational frequencies have
 reached a relative accuracy of a few parts in $10^{-10}$
 \cite{korobov2008},
 and the influence of external perturbing
 and exciting electromagnetic fields, i.e. all systematics,
 can also be treated accurately ab initio.
 Comparison of theoretical and experimental transition
 frequencies in HD$^+$ can potentially lead to determining the electron to
 proton/deuteron mass ratio and the Rydberg constant
 in an alternative way.
 For these
 purposes, an experimental transition frequency accuracy in the
 range $1\cdot10^{-10}$ to $1\cdot10^{-14}$ must be attained. The
 requirement is even more stringent, at the $10^{-16}$ level, if the
 system is to be used to test the time independence of the mass
 ratio or Lorentz Invariance.
 Experimentally, HD$^+$ can be cooled
 to tens of mK by sympathetic cooling in an ion trap \cite{Blythe} and
 rotationally cooled \cite{tschneider}. One-photon laser spectroscopy of
 ro-vibrational transitions has been performed and one transition
 frequency has been determined with a relative uncertainty at the
 $2\cdot10^{-9}$ level \cite{schiller08}, in
 agreement with the QED calculation \cite{korobov2008}.

Here we present the results of a thorough  study of the magnetic field effects on the
 radiofrequency, rotational (THz) and ro-vibrational transitions in
 HD$^+$. 
 Both one- and two-photon transitions are of
 interest for precision spectroscopy. Earlier, Karr et al.
 \cite{jphkarr} had evaluated the two-photon transitions strengths in
 HD$^+$ in the spinless particle approximation.
 An analysis of a particular
 ro-vibrational transition at moderate spectral resolution has
 been reported in Ref.\cite{schiller08}.

 {\em Theoretical approach.} The Hamiltonian $H$ of the HD$^+$ ion
 in an external magnetic field has
 the form $H=H^{\rm NR}+V^{\rm diag}+
 V^{\rm spin}+ V^{\rm mag}$, where $H^{\rm NR}$ is the
 non-relativistic 3-body Hamiltonian 
 and the correction
 terms collect the spin-independent interactions, the spin
 interactions (cf. Ref.~\cite{PRL06,LNP07})
 and the external magnetic field interaction terms,
 respectively. In the leading order approximation
 $V^{\rm mag}=
 -{\mathbf B}\cdot\sum_i (eZ_i/2M_ic)({\mathbf L}_i+(\mu_i/s_i)
 {\mathbf S}_i)$,
 where the summation is over the constituents of HD$^+$
 $(i=p,d,e)$,
 $M_i$, $Z_i$ and $\mu_i$ are the mass,
 the electric charge (in units $e$) and
 the magnetic moment (in units $e\hbar/2M_ic$)
 of particle ``$i$'',
 $s_i$ is the size of its spin ($1/2$ for $i=p,e$
 and $1$ for $i=d$),
 ${\mathbf L}_i$ and
 ${\mathbf S}_i$ are the
 orbital and spin angular momentum operators in the center-of-mass-frame,
 and ${\mathbf B}$ is the external magnetic field.
 Neglecting the higher order corrections to $V^{\rm mag}$
 (cf. Ref.~\cite{hegg}) is justified for magnetic fields $B$ below the
 threshold $B_{thr}\sim 10^2$ G for which the contribution of
 $V^{\rm mag}$, increasing with $B$, reaches the order of
 magnitude of the hyperfine energy, since the
 relativistic corrections to $V^{\rm mag}$
 are smaller than the theoretical uncertainty of the
 hyperfine energy levels of Ref.~\cite{PRL06}.

 The spin structure of the ro-vibrational state $(v,L)$,
 where $v$ and $L$ are the vibrational and
 total orbital momentum quantum numbers, is calculated in first
 order of perturbation theory using an effective
 Hamiltonian $V^{\rm eff}$
 that is obtained by averaging
 $V^{\rm spin}+V^{\rm mag}$
 over the spatial degrees of freedom.
 Compared to $H_{\rm eff}$ (the
 HD$^+$ effective spin Hamiltonian of Ref.~\cite{PRL06}),
 $V^{\rm eff}$ includes 4 additional terms,
 originating from $V^{\rm mag}$:
 $V^{\rm eff}=H_{\rm eff}+
 E_{10}({\mathbf L}\cdot{\mathbf B})+
 E_{11}({\mathbf S}_p\cdot{\mathbf B})+
 E_{12}({\mathbf S}_d\cdot{\mathbf B})+
 E_{13}({\mathbf S}_e\cdot{\mathbf B})$.
 In the adopted leading order approximation
 $E_{11}=-4.2577$ kHz/G, $E_{12}=-0.6536$ kHz/G and
 $E_{13}=2.8025$ MHz/G are expressed only in terms of
 the mass and magnetic moments of the particles.
 The values of $E_{10}$ were
 calculated using the
 variational non-relativistic wave functions of HD$^+$ of
 Ref.~\cite{DDBkorobov}, thus
 improving on the accuracy of the first such calculations \cite{hegg}.

 The matrix of $V^{\rm eff}$ is evaluated in the basis set
 of vectors
 with definite values of the
 squared angular momenta ${\mathbf F}={\mathbf S}_p+{\mathbf S}_e$,
 ${\mathbf S}={\mathbf F}+{\mathbf S}_d$,
 ${\mathbf J}={\mathbf S}+{\mathbf L}$ and $z$-axis projection $J_z$
 of 
 ${\mathbf J}$.
 Except for the ``stretched'' states with the ``extreme'' values
 $F\!=\!1$, $S\!=\!2$, $J\!=\!L+2$ and $J_z\!=\!\pm J$,
 $J_z$ is the only exact quantum number; labeling the eigenstates
 of $V^{\rm eff}$ requires an additional index $n$.
 The corresponding eigenvalues $\Delta E^{vLnJ_z}$
 represent the energy levels of HD$^+$,
 defined relative to the ``spinless'' energies $E^{vL}$,
 calculated as
 eigenvalues of $H^{\rm NR}+V^{\rm diag}$.
 Since the hyperfine
 spectrum is only deformed but not rearranged by magnetic fields
 $B$ below a few G, we take for $n$ the set of quantum
 numbers $(F, S, J)$ labelling the hyperfine states at $B\!=\!0$.
 %

 The hyperfine states of HD$^+$ are split into sublevels distinguished
 with the quantum number $J_z$.
 The Zeeman shift may be approximated with the
 quadratic form
 \begin{equation}
 \Delta E^{vLnJ_z}(B)-\Delta E^{vLnJ_z}(0)\approx h^{vLn} J_z  B+q^{vLnJ_z} B^2\,.
 \label{quadradic}
 \end{equation}
 The numerical values of $h^{vLn}$ and  $q^{vLn0}$,
 calculated by the least square method and providing
 relative uncertainty below $10^{-6}$ for $B<1$ G
 are given in Table~\ref{h-lin}.
 Eq.~(\ref{quadradic}) may be used to evaluate the Zeeman shift of
 transition frequencies
 and choose the less
 sensitive ones as candidates for precision spectroscopy.
 Stretched states are a special case: the Zeeman shift is {\em
 strictly} linear, with
 $h^{vLn}=(L\,E_{10}+E_{12}+(E_{11}+E_{13})/2)/(L+2)$.

\begin{figure}[ht]
 \includegraphics[width=.5\textwidth]{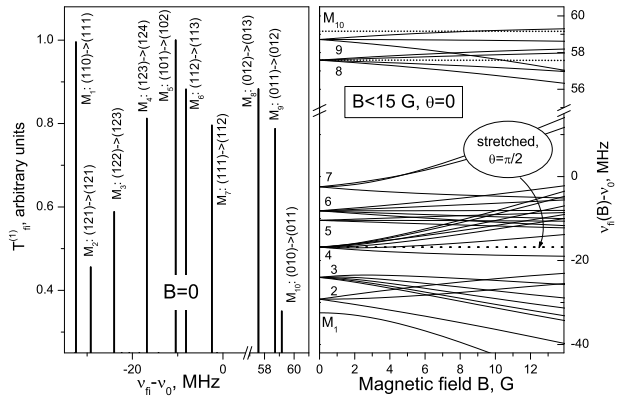}
 \caption{Left: Strength $T^{(1)}_{fi}$
 of ten favored hyperfine components $M_1,\ldots,M_{10}$ of the
 $(0,1)\rightarrow(4,2)$ one-photon transition line at $B\!=\!0$
 (no $\theta$-dependence),
 labeled with the quantum numbers $(FSJ)$ of the initial and final states.
Right: Frequencies of the $\Delta J_z\!=\!0$ components of
 $M_1,\ldots,M_{10}$ as function of the magnetic field $B$ for $\theta\!=\!0$.
 The $J_z\!=\!0\rightarrow J'_z\!=\!0$ components of $M_8$ and $M_{10}$
 (short dotted lines) undergo a very small quadratic shift.
 The frequencies of the transitions between
 stretched states in transversal field (dotted lines) have a very small
 linear shift; their mean is independent of $B$.
}
 \label{XXX3}
 \end{figure}

 {\em Transitions.}
 The probability per unit time
 for an electric dipole
 transition between a lower $|i\rangle\!\equiv\!|vLnJ_z\rangle$
 and upper $\langle f|
 \!\equiv\!\langle v'L'n'J'_z|$ states,
 stimulated by an oscillating electric field
 ${\mathbf E}(t)={\mathbf E}\,\cos 2\pi\nu t$
 of frequency $\nu$, is (in units $h=1$)
 $\pi^2
 \delta(\nu\!-\!\nu_{fi})
 |\langle f|W|i\rangle|^2$, where
 $\nu_{fi}\!=\!\nu_0\!+\!\Delta E^{v'L'n'J'_z}\!-\!
  \Delta E^{vLnJ_z}$,
 $\nu_0\!=\!E^{v'L'}\!-\!E^{vL}$ is referred to as
 ``central frequency'',\,
 $W\!=\!-{\mathbf E}.{\mathbf d}$ is the interaction with the
 electric field, and ${\mathbf d}$ is the
 electric dipole moment operator.
 The matrix elements
 $\langle f|W|i\rangle$ are expressed in terms of
 the reduced matrix elements of
 ${\mathbf d}$ which, in the non-relativistic approximation,
 do not depend of the spin quantum numbers: $d_{fi}\equiv
 d_{v'L',vL}$.
 Accurate numerical values of $d_{v'L',vL}$ have been
 calculated with the variational Coulomb wave functions of
 Ref.~\cite{DDBkorobov};
 they agree with earlier results of
 Ref.~\cite{bunker} at the $10^{-3}$ level.
 Assuming statistical population of the hyperfine states,
 the observable spectrum of the
 transition $(v,L)\!\rightarrow\!(v',L')$
 can be put in the form
 $|{\mathbf E}|^2
 \sum_{f,i}\delta(\nu-\nu_{fi})
 \,T^{(1)}_{fi}(\theta)$,
 where $\theta$ is the angle between ${\mathbf B}$
 and ${\mathbf E}$,
 the sum is over all pairs of states $(i,f)$
 belonging to the hyperfine structure of the initial and
 final ro-vibrational states.
 The number of hyperfine lines may exceed $10^3$, but most of them are weak. For
 magnetic fields $B$ below a few G the spectrum
 is dominated 
 by the ``favored'' components corresponding to transitions between states
 that satisfy $\Delta S\!=\!\Delta F\!=\!0$; the frequencies of the favored
 transitions lie in a
 band of width $\sim100$ MHz around $\nu_0$.
 Fig.~\ref{XXX3} illustrates the Zeeman structure of an overtone ro-vibrational
 transition ($(0,1)\rightarrow(4,2)$), to be discussed further below.

 Of interest also are the radio-frequency (RF) magnetic dipole (M1) transitions
 between spin sub-states of the
 same $(vL)$ state, stimulated by an oscillating magnetic field ${\mathbf B}'\cos 2\pi\nu t$.
 The frequencies $\nu_{fi}$ of the transitions with $\Delta F=\pm1$ are in the
 range $700<\nu_{fi}<1100$ MHz, while $\nu_{fi}<200$ MHz for $\Delta F=0$.
 The explicit expressions of the RF transition probabilities show that they
 are mainly due to the $({\mathbf B}'\cdot{\mathbf
 S}_e)$ term in $V^{\rm mag}$.

The two-photon spectrum may similarly be put in the form
 $|{\mathbf E}|^4\sum_{f,i}\delta(\nu\!-\!\nu_{fi}/2)\,
 T^{(2)}_{fi}(\theta)$.
 In transverse polarization ($\theta\!=\!\pi/2$) each favored
 hyperfine component
 splits into 3 components with $\Delta J_z=J'_z-J_z=-2,0,2$ with a typical separation
 of the order of $\sim 10^2$ kHz at $B=0.5$ G; each of these components
 acquires an additional super-fine
 structure with separations in the $10^1$ kHz range.
 In a longitudinal magnetic field ($\theta=0$)
 $\Delta J_z=0$, and the lines are ``super-fine'' split only.

Transitions between stretched states are allowed by selection rules, both
as one-photon pure rotational ($v=v'$) or ro-vibrational as well as two-photon transitions.

 {\em Discussion.}
 At low $B$, the Zeeman shift of the frequency of a single Zeeman component may be estimated
 using Eq.~(\ref{quadradic}): $\delta\nu_{fi}(B)\approx
(h'J'_z-hJ_z)B+ (q'-q)B^2$.
 The systematic uncertainty due to the external
 magnetic field may be
 minimized by selecting transitions with minimal $\delta\nu_{fi}(B)$.
We describe a few concrete transitions, for different cases of
assumed experimental spectral resolution.

{\em 1: Resolved hyperfine structure, unresolved Zeeman
subcomponents.} The level of spectroscopic resolution needed for
improved measurements of mass ratios is about 10 kHz for an
overtone vibrational transition and therefore requires
experimental resolution of the hyperfine structure (with typical
spacing of several MHz, see Fig.~\ref{XXX3}), but not necessarily
of the Zeeman sublevels.

\begin{figure}[ht]
 \includegraphics[width=.5\textwidth]{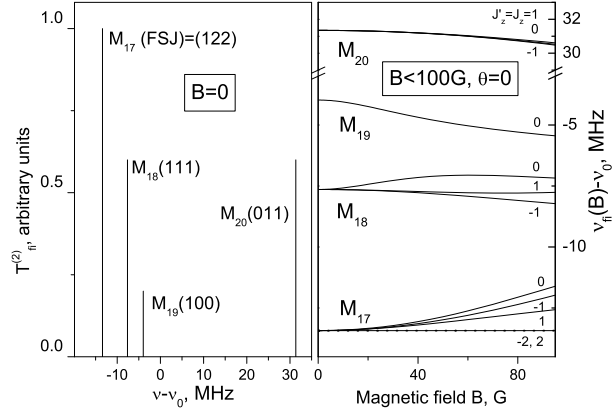}
 \caption{Left: strength $T^{(2)}_{fi}$
 of the favored components of the
 $(0,0)\rightarrow(2,0)$ two-photon transition line at $B\!=\!0$,
 labelled with the quantum numbers $(FSJ)\!=\!(F'S'J')$.
 Right: Zeeman shift of the $\Delta J_z\!=\!0$ components
 in external magnetic field $B$. The spectrum is independent of $\theta$.
 The transition frequencies between stretched states
 ($J_z\!=\!\pm2$ components of $M_{17}$, dotted line) are
 independent of $B$.
 }
 \label{2gamma-ground}
 \end{figure}

 Among one-photon transition, a favorable case is the triplet of magnetic subcomponents
 $|vL(FSJ)J_z\rangle=|01(010)0\rangle\rightarrow |42(011)J_z'\rangle$.
 The $J_z\!=\!0\rightarrow J_z'\!=\!0$ subcomponent (parallel polarization) has
 a very small quadratic Zeeman shift (-18 Hz at 1 G), while the
 $ J_z\!=\!0\rightarrow J_z'\!=\!\pm1$ subcomponents (observable in
 perpendicular polarization) have dominant linear Zeeman shifts of
 34.3 and $-$36.3 kHz, respectively, with an intensity-weighted
 shift of only 290 Hz at 1 G ($1.3\cdot10^{-12}$ in relative units). If the magnetic
 field direction is not optimal, the latter value sets the scale of the Zeeman
 systematic effect.
 A two-photon transition candidate is the $\theta$-insensitive
$|00(011)J_z\rangle\rightarrow |20(011)J_z\rangle$ triplet (line
$M_{20}$ in Fig.~\ref{2gamma-ground}) with a weighed shift of
$-$95 Hz at 1 G $(-8\cdot10^{-13})$.
 As fields below 1 G are feasible,
 such shift values will be well below the
current and near-future theoretical uncertainties and thus will
not be a limitation in comparisons of theoretical and experimental
results.

{\em 2: Resolved Zeeman subcomponents:} the preferred experimental
situation, in which the
 transition frequencies of individual subcomponents can be measured.
 Examples of one-photon transitions with very low magnetic field
 sensitivity include transitions with $J'_z\!=\!J_z\!=\!0$ with
 eliminated linear and suppressed quadratic dependence on $B$: 
 $|vL(FSJ)J_z\rangle\!=\!|01(011)0\rangle\!\rightarrow\!|02(012)0\rangle$
 (pure rotational),
 $|01(112)0\rangle\!\rightarrow\!|32(113)0\rangle$,
 $|01(010)0\rangle\!\rightarrow\!|42(011)0\rangle$,
 $|01(012)0\rangle\!\rightarrow\!|42(013)0\rangle$,
 $|02(012)0\rangle\!\rightarrow\!|43(013)0\rangle$, with sensitivities
 $\delta\nu(1\mathrm G)/\nu_0\!\approx\!(3,0.1,-0.08,-0.2,-0.2)\cdot10^{-12}$,
 respectively. 
 The 
 sensitivity may also be weak in transitions
 with $\Delta J_z\ne0$, e.g. in
 $|02(011){-1}\rangle\!\rightarrow\!|43(012){-2}\rangle$ with
 $|\delta\nu(B)/\nu_0|<8\cdot10^{-13}$ for $B<1\,$G.

One-photon pure rotational or ro-vibrational transitions between stretched
 states occur as a doublet with purely linear magnetic shift:
 $\delta\nu_{fi}(B)=\pm(L'\,E'_{10}-L\,E_{10})B$ for
 $J_z=\pm( L+2)\rightarrow J_z'=\pm( L'+2)$, where $L,E_{10}$
 $(L',E'_{10})$ refer to the initial (final) state, and $|L-L'|=1$. This is about
 $\pm 0.5\,$kHz for $B=1$ G. The mean of the frequencies is
 strictly {\em independent} of $B$ in the adopted approximation.
It should be possible to measure each doublet component
independently and then compute the average, as in atomic optical
clocks.

 Fig.~\ref{XXX3} gives an example of a one-photon transition,
 $(0,1)\rightarrow(4,2)$ at $\lambda\approx1.4\,$$\mu$m.
 The hyperfine spectrum includes 18 favored lines of which 10 are plotted.
 While most of them
 are quite sensitive to the external magnetic field strength,
 the $J_z\!=\!0\rightarrow J_z'\!=\!0$ subcomponents of $M_{8}$ and $M_{10}$
 are weakly sensitive to $B$.
 Line $M_{10}$
 also contains the moderately sensitive doublet mentioned above,
 and  $M_4$ contains the stretched states transitions observable
 at perpendicular polarization only.

Two-photon transitions between stretched states are of
metrological interest. In transitions with $L'=L\pm2$ the doublet
splitting is of order $\pm 10\,$Hz/G, since $E_{01}$ varies weakly
with the vibrational quantum number, and the mean shift is zero.
Again, it should be possible to measure each doublet component
independently and then compute the mean. For two-photon
transitions with $L'\!=\!L\!=\!0$ there is no splitting and no
shift. Favourable two-photon transitions are
$(0,0)\rightarrow(2,0)$ at $2\times5.362\,\mu$m and
$(0,1)\rightarrow(2,1)$ at $2\times5.368\,\mu$m, which have
intermediate states whose detuning is not too large and therefore
allow a reasonable excitation strength for realistic laser power
from quantum cascade lasers. In addition, the
$(0,0)\rightarrow(2,0)$ transition has a particularly simple
spectrum, see Fig. \ref{2gamma-ground}, since the $L\!=\!0$ states
have only four levels and the two-photon transition selection
rules \cite{bcagnac} allow only
 $\Delta J\!=\!\Delta J_z\!=\!0$
 transitions. This gives a relatively strong weight to the
stretched states transition relative to the sum over all
transitions (which are all equally strong). Its frequency is given
by a simple expression, see below.


 {\em 3. Comparison of ab-initio theory and experiment.}
 When spectroscopy of HD$^+$
 is pursued with the goal of comparison with QED calculations and
 for a determination of the particle mass ratio, it may
 be important to determine the central frequency $\nu_0$
 since in the adopted approximation,
 unlike the spin corrections, $\nu_0$
 includes the contribution of QED corrections of order
 $O(m\alpha^5)$ and $O(m\alpha^5\log\alpha)$.
 We propose to combine the results of optical
 pectroscopy between
 vibrational levels $(v,L)$ and $(v',L')$ with the results of RF spectroscopy
 within each of these levels to extract $\nu_0$.
 Indeed, the coefficients of the effective spin Hamiltonian $E_i$
 and the spin corrections to $E^{vL}$, can be expressed in terms of
 the frequencies of RF transitions within the hyperfine structure of the $(v,L)$
 state (and similar for the $(v',L')$ state).
 Since the number $n_h$ of linearly independent hyperfine transition
 frequencies 
 exceeds
 the number $N_e$ of non-vanishing $E_i$, 
 the above relations can be resolved
 for $E_i$ only if the radio-frequencies
 satisfy $N_h-N_e$ compatibility relations, that may
 serve to test the experimental accuracy.
 The value of $\nu_0$ is then expressed in terms of
 spectroscopic data and may be used to test the predictions of
 few-body bound state QED in next-to-leading orders in $\alpha$.
 This approach is particularly simple in states with $L\!=\!0$, since the
 only non-vanishing coefficients are $E_4$ and $E_5$.
 Denote the frequencies of the
 $J_z\!=\!0\rightarrow J'_z\!=\!0$ components at $B=0$
 of three RF transitions in the $(v,L=0)$ state as follows:
 $\nu_1=\nu_{v0(111)0,v0(100)0}$,
 $\nu_2=\nu_{v0(122)0,v0(111)0}$
 $\nu_3=\nu_{v0(100)0,v0(011)0}$, and similarly, by
 $\nu'_1\ldots,\nu'_3$, the corresponding frequencies
 in the $(v',0)$ state.
Diagonalizing the hyperfine hamiltonian, we obtain $E_4=\nu_2+\nu_3$,
$E_5=2\,(\nu_1+\nu_2)/3$, with
 compatibility relation $\nu_2(\nu_1+\nu_2+\nu_3)=2\nu_1\nu_3$
 (and similar for the $(v',0)$ state).
The spin energy of the stretched states is
$\Delta E^{v0(122)2}=E_4/4+E_5/2=\nu_1/3 + 7\nu_2/12 +\nu_3/4$.
The transition  frequency  between stretched states is
 $\nu_{v0(122)2,v'0(122)2}=\nu_0+\Delta\nu$, $\Delta\nu=
 1/3\,(\nu'_1- \nu_1) + 7/12\,(\nu'_2-
 \nu_2)+(\nu'_3-\nu_3)/4$. Thus, $\nu_0$ could be obtained  by measuring
 one vibrational transition, and six RF transitions. The sensitivity to
 magnetic field comes only from $\Delta\nu$, since
 the Zeeman shift of $\nu_{v0(122)2,v'0(122)2}$ is strictly zero in
 the adopted approximation.  For example, $\Delta\nu$ is shifted by
 only 55 Hz ($5\cdot10^{-13}$) in the case $v=0, v'=2$
 at $B=1$ G.
 This shift value could be taken as a conservative Zeeman uncertainty for
 $\nu_0$, but it could be further reduced
by correcting for the magnetic field in the trap; the field could be
determined e.g. from a measurement of the Zeeman splitting of an
appropriate magnetic-sensitive transition and using the
 theoretical magnetic sensitivities.

 {\em 4. Spectroscopic techniques.}
In order to achieve the spectroscopic resolutions discussed, a
Doppler-free technique is required, and two methods appear
suitable. Two-photon spectroscopy with counter-propagating beams
of an ensemble of cooled, but not strongly confined molecular ions
strongly suppresses the first-order Doppler broadening
\cite{jcbergquist}. Second is  quantum logic spectroscopy
\cite{trosenband} on a single HD$^+$ ion. For both methods,
techniques for strongly populating  the lower spectroscopic state
will be helpful \cite{tschneider}.

{\em  Summary.} By evaluating the Zeeman effect for all
experimentally relevant spectroscopies of HD$^+$ we have shown
 that it will not be a limiting factor for the experimental accuracy if
 appropriate transitions and spectroscopic technique are selected.
 This is of particular relevance to the possible use of HD$^+$ for setting
 limits to a
 hypothetical time-dependence of particle mass ratios, where a
 relative accuracy of $10^{-16}$ is desired.
 For the determination of the ratio of electron to reduced nuclear
 mass by comparison of {\em ab initio} QED calculations
 and experimental frequencies, it is advantageous to determine the central
 transition frequencies $\nu_0$
 since the uncertainties related to particle magnetic
 moments and nuclear electromagnetic structure will be significantly
 suppressed. We have shown how this can be achieved by combining
 two-photon and RF spectroscopy of states with low magnetic
 sensitivity, the latter feature not being affected by the higher-order
 relativistic mass and anisotropy corrections.

 Other systematic shifts, such as light, 
 blackbody radiation 
 and Stark shifts are currently being analyzed.


{\em Acknowledgments} The work was partly supported by Grant
02-288 of the Bulgarian Science Fund (D.B.) and the Russian
Foundation for Basic Research under Grant No. 08-02-00341
(V.I.K.). D.B. and V.I.K also acknowledge the support from a JINR
grant for the joint research of BLTP-JINR and INRNE-BAS.
 The work of S.S. was supported by DFG project Schi 431/11-1.

\vfill\eject

 \newpage
 \setlength{\topmargin}{-3.5cm}
 \thispagestyle{empty}
\begin{table}[ht]
 \begin{center}
 \caption{Numerical values of the coefficients
 $h^{vLn}$ (upper line, in kHz/G)
 and
 $q^{vLnJ_z}, J_z=0$ (lower line, in kHz/G$^2$)
 in the weighed least-square approximation (1) of the dependence of
 the Zeeman corrections to the hyperfine energy levels on the
 magnetic field $B$ for the ro-vibrational states of the
 HD$^+$ ion with $L\le4$ and $v\le4$.
 The relative inaccuracy of the approximation does not exceed $5.10^{-7}$ for $B<1$ G,
 except for the $F=1$ components of the states with $L=1$ where accuracy better than
 $10^{-5}$ can only be achieved with higher degree polynomial approximation.}
 \label{h-lin}
 \begin{tabular}{l@{\hspace{2mm}}rrr@{\hspace{3mm}}r@{\hspace{3mm}}rrr@{\hspace{3mm}}rrrrr}
  \hline\hline
  \vspace*{2pt}
  &
  \multicolumn{3}{c}{{\vrule height 12pt width 0pt}$(F,S)=(0,1)$} &
  \multicolumn{1}{c}{$(1,0)$} &
  \multicolumn{3}{c}{$(1,1)$} &
  \multicolumn{5}{c}{$(1,2)$} \\[2pt]
  \hline
  {\vrule height 12pt width 0pt}
  $vL$ &
  \multicolumn{1}{c}{$J\!=\!L\!+\!1$} &
  \multicolumn{1}{c}{$L$} &
  \multicolumn{1}{c}{$L\!-\!1$} &
  \multicolumn{1}{c}{$L$} &
  \multicolumn{1}{c}{$L\!+\!1$} &
  \multicolumn{1}{c}{$L$} &
  \multicolumn{1}{c}{$L\!-\!1$} &
  \multicolumn{1}{c}{$L\!+\!2$} &
  \multicolumn{1}{c}{$L\!+\!1$} &
  \multicolumn{1}{c}{$L$} &
  \multicolumn{1}{c}{$L\!-\!1$} &
  \multicolumn{1}{c}{$L\!-\!2$} \\[2pt]
  \hline
 00 &     -218.820 &              &              &        0.000 &      917.397 &              &              &      699.233 &              &              &              &             \\
    &       -2.074 &              &              &      -17.657 &       14.784 &              &              &        4.947 &              &              &              &             \\
 10 &     -218.835 &              &              &        0.000 &      917.411 &              &              &      699.233 &              &              &              &             \\
    &       -2.122 &              &              &      -18.071 &       15.130 &              &              &        5.063 &              &              &              &             \\
 20 &     -218.851 &              &              &        0.000 &      917.428 &              &              &      699.233 &              &              &              &             \\
    &       -2.170 &              &              &      -18.475 &       15.468 &              &              &        5.176 &              &              &              &             \\
 30 &     -218.871 &              &              &        0.000 &      917.448 &              &              &      699.233 &              &              &              &             \\
    &       -2.217 &              &              &      -18.869 &       15.798 &              &              &        5.287 &              &              &              &             \\
 40 &     -218.894 &              &              &        0.000 &      917.470 &              &              &      699.233 &              &              &              &             \\
    &       -2.262 &              &              &      -19.251 &       16.119 &              &              &        5.394 &              &              &              &             \\
 01 &     -129.368 &     -148.704 &        0.000 &     -473.172 &      466.638 &      821.918 &        0.000 &      465.376 &      595.729 &     1197.227 &              &             \\
    &       -4.988 &       -2.940 &        1.743 &      -12.606 &        8.611 &      -95.776 &       92.665 &      -95.865 &      106.989 &        2.167 &              &             \\
 11 &     -128.783 &     -147.464 &        0.000 &     -459.754 &      466.337 &      809.370 &        0.000 &      465.291 &      595.691 &     1195.121 &              &             \\
    &       -5.222 &       -3.121 &        2.015 &      -13.127 &        9.133 &      -99.725 &       96.462 &     -104.234 &      115.802 &        2.019 &              &             \\
 21 &     -128.183 &     -146.200 &        0.000 &     -445.608 &      466.036 &      796.288 &        0.000 &      465.198 &      595.666 &     1192.796 &              &             \\
    &       -5.470 &       -3.317 &        2.318 &      -13.657 &        9.694 &     -103.890 &      100.450 &     -113.351 &      125.380 &        1.843 &              &             \\
 31 &     -127.569 &     -144.909 &        0.000 &     -430.712 &      465.734 &      782.670 &        0.000 &      465.096 &      595.651 &     1190.231 &              &             \\
    &       -5.735 &       -3.529 &        2.658 &      -14.193 &       10.301 &     -108.308 &      104.657 &     -123.282 &      135.795 &        1.635 &              &             \\
 41 &     -126.934 &     -143.585 &        0.000 &     -415.040 &      465.434 &      768.518 &        0.000 &      464.990 &      595.642 &     1187.400 &              &             \\
    &       -6.018 &       -3.762 &        3.042 &      -14.736 &       10.961 &     -113.021 &      109.119 &     -134.081 &      147.104 &        1.391 &              &             \\
 02 &     -100.293 &      -80.709 &       24.335 &     -364.548 &      307.231 &      442.688 &     -650.959 &      349.337 &      374.520 &      466.911 &      624.618 &        0.000\\
    &       -4.198 &       -2.932 &        0.904 &      -11.457 &        7.212 &       32.570 &      -35.737 &       37.105 &      -17.480 &        7.061 &       25.264 &      -38.314\\
 12 &      -99.490 &      -79.360 &       27.011 &     -357.541 &      307.093 &      437.197 &     -646.350 &      349.339 &      373.864 &      464.072 &      617.290 &        0.000\\
    &       -4.387 &       -3.058 &        1.076 &      -11.972 &        7.650 &       35.119 &      -38.324 &       38.907 &      -18.322 &        7.352 &       26.707 &      -40.748\\
 22 &      -98.667 &      -77.979 &       29.748 &     -350.004 &      306.960 &      431.260 &     -641.463 &      349.342 &      373.184 &      461.119 &      609.615 &        0.000\\
    &       -4.587 &       -3.190 &        1.268 &      -12.503 &        8.122 &       37.889 &      -41.147 &       40.830 &      -19.226 &        7.658 &       28.256 &      -43.371\\
 32 &      -97.821 &      -76.563 &       32.551 &     -341.885 &      306.832 &      424.836 &     -636.285 &      349.345 &      372.478 &      458.041 &      601.573 &        0.000\\
    &       -4.799 &       -3.329 &        1.484 &      -13.047 &        8.633 &       40.913 &      -44.244 &       42.891 &      -20.201 &        7.981 &       29.928 &      -46.210\\
 42 &      -96.945 &      -75.101 &       35.443 &     -333.120 &      306.715 &      417.886 &     -630.789 &      349.353 &      371.747 &      454.830 &      593.142 &        0.000\\
    &       -5.026 &       -3.478 &        1.727 &      -13.606 &        9.190 &       44.230 &      -47.658 &       45.113 &      -21.260 &        8.326 &       31.739 &      -49.298\\
 03 &      -85.696 &      -63.510 &        2.629 &     -296.799 &      225.824 &      311.507 &     -423.624 &      279.364 &      278.044 &      279.838 &      186.045 &     -700.466\\
    &       -3.705 &       -2.584 &        0.060 &       -9.894 &        5.759 &       15.852 &      -18.897 &       20.502 &        0.673 &        6.205 &        8.318 &      -22.289\\
 13 &      -84.790 &      -62.138 &        4.836 &     -292.840 &      225.798 &      308.857 &     -420.847 &      279.367 &      277.175 &      277.177 &      181.073 &     -700.471\\
    &       -3.865 &       -2.675 &        0.167 &      -10.377 &        6.114 &       17.059 &      -20.106 &       21.586 &        0.636 &        6.430 &        8.663 &      -23.633\\
 23 &      -83.861 &      -60.733 &        7.093 &     -288.510 &      225.782 &      305.901 &     -417.911 &      279.371 &      276.272 &      274.419 &      175.891 &     -700.478\\
    &       -4.033 &       -2.770 &        0.289 &      -10.879 &        6.499 &       18.360 &      -21.414 &       22.745 &        0.590 &        6.664 &        9.027 &      -25.078\\
 33 &      -82.904 &      -59.289 &        9.408 &     -283.760 &      225.777 &      302.597 &     -414.802 &      279.375 &      275.335 &      271.554 &      170.482 &     -700.484\\
    &       -4.210 &       -2.867 &        0.427 &      -11.402 &        6.917 &       19.767 &      -22.835 &       23.991 &        0.532 &        6.908 &        9.412 &      -26.640\\
 43 &      -81.913 &      -57.797 &       11.796 &     -278.525 &      225.789 &      298.905 &     -411.500 &      279.385 &      274.363 &      268.576 &      164.836 &     -700.477\\
    &       -4.399 &       -2.970 &        0.586 &      -11.947 &        7.374 &       21.296 &      -24.388 &       25.336 &        0.460 &        7.163 &        9.824 &      -28.336\\
 04 &      -76.724 &      -56.329 &       -8.945 &     -248.212 &      176.961 &      237.613 &     -317.629 &      232.709 &      224.638 &      204.655 &       91.646 &     -467.118\\
    &       -3.387 &       -2.407 &       -0.434 &       -8.642 &        4.784 &       10.241 &      -13.379 &       14.907 &        5.129 &        5.664 &        3.631 &      -16.106\\
 14 &      -75.767 &      -54.959 &       -6.945 &     -245.779 &      176.976 &      236.409 &     -315.688 &      232.713 &      223.678 &      202.075 &       87.721 &     -467.114\\
    &       -3.527 &       -2.483 &       -0.363 &       -9.083 &        5.079 &       11.062 &      -14.192 &       15.714 &        5.315 &        5.842 &        3.694 &      -17.059\\
 24 &      -74.783 &      -53.554 &       -4.898 &     -243.092 &      177.005 &      235.011 &     -313.627 &      232.717 &      222.678 &      199.401 &       83.629 &     -467.110\\
    &       -3.674 &       -2.560 &       -0.280 &       -9.546 &        5.399 &       11.949 &      -15.067 &       16.577 &        5.507 &        6.025 &        3.754 &      -18.084\\
 34 &      -73.770 &      -52.109 &       -2.797 &     -240.113 &      177.049 &      233.388 &     -311.433 &      232.722 &      221.635 &      196.626 &       79.356 &     -467.104\\
    &       -3.828 &       -2.639 &       -0.184 &      -10.032 &        5.747 &       12.908 &      -16.014 &       17.506 &        5.704 &        6.212 &        3.812 &      -19.190\\
 44 &      -72.720 &      -50.616 &       -0.632 &     -236.790 &      177.114 &      231.504 &     -309.090 &      232.731 &      220.549 &      193.743 &       74.889 &     -467.090\\
    &       -3.993 &       -2.720 &       -0.072 &      -10.545 &        6.130 &       13.949 &      -17.046 &       18.510 &        5.907 &        6.404 &        3.867 &      -20.391
 \end{tabular}
 \end{center}
 \end{table}

\end{document}